\documentclass[pdflatex,sn-mathphys-num]{sn-jnl}

\usepackage{graphicx}%
\usepackage{multirow}%
\usepackage{amsmath,amssymb,amsfonts}%
\usepackage{amsthm}%
\usepackage{mathrsfs}%
\usepackage[title]{appendix}%
\usepackage{xcolor}%
\usepackage{textcomp}%
\usepackage{manyfoot}%
\usepackage{booktabs}%
\usepackage{algorithm}%
\usepackage{algorithmicx}%
\usepackage{algpseudocode}%
\usepackage{listings}%

\usepackage{soul} 
\usepackage{lscape}

\theoremstyle{thmstyleone}%
\theoremstyle{thmstyletwo}%

\theoremstyle{thmstylethree}%

\begin{document}

\title[Article Title]{An Empirical Evaluation of Modern MLOps Frameworks}


\author[1]{\fnm{Jon} \sur{Marcos-Mercad\'e}}\email{jmarcos021@ehu.eus}

\author[1]{\fnm{Unai} \sur{Lopez-Novoa}}\email{unai.lopez@ehu.eus}

\author[1]{\fnm{Mikel} \sur{Ega\~na Aranguren}}\email{mikel.egana@ehu.eus}

\affil[1]{\orgdiv{Department of Computer Languages and Systems}, \orgname{University of the Basque Country UPV/EHU}, \orgaddress{\postcode{48013}, \city{Bilbao}, \country{Spain}}}


\abstract{Given the increasing adoption of AI solutions in professional environments, it is necessary for developers to be able to make informed decisions about the current tool landscape. This work empirically evaluates various MLOps (Machine Learning Operations) tools to facilitate the management of the ML model lifecycle: MLflow, Metaflow, Apache Airflow, and Kubeflow Pipelines. The tools are evaluated by assessing the criteria of Ease of installation, Configuration flexibility, Interoperability, Code instrumentation complexity, Result interpretability, and Documentation when implementing two common ML scenarios: Digit classifier with MNIST and Sentiment classifier with IMDB and BERT. The evaluation is completed by providing weighted results that lead to practical conclusions on which tools are best suited for different scenarios. }




\maketitle

\section{Introduction}\label{sec1}

In recent years, Artificial Intelligence (AI) has evolved from a primarily academic discipline to a pervasive technology adopted by individuals and organizations across all sectors. As highlighted by recent market analyses \cite{ref-90}, the number of daily AI users has grown dramatically—from 116 million in 2020 to 314 million in 2024, with projections estimating nearly 730 million daily users by 2030. This rapid adoption is driven by the massive availability of data, increased computational power, and the development of more efficient algorithms, enabling organizations to integrate Machine Learning (ML) solutions into products, processes, and services to enhance decision-making, optimize operations, and personalize user experiences.

Concurrently, the DevOps paradigm has gained prominence in software engineering, promoting automation, collaboration between development and operations teams, and continuous delivery. DevOps practices have enabled organizations to shorten development cycles, improve product quality, and deploy software more securely and at scale \cite{ref-91}.

However, integrating AI and DevOps presents unique technical and organizational challenges not encountered in traditional software development. Key issues include managing data and model versioning, addressing performance degradation over time (data drift), ensuring experiment reproducibility, and automating continuous retraining and validation. To address these challenges, the Shift Left approach has emerged, advocating for the early integration of critical tasks such as validation, testing, and quality control within the development lifecycle \cite{ref-60}. In the context of ML, Shift Left emphasizes early attention to data quality, versioning, experiment traceability, metrics definition, and continuous validation, thereby facilitating collaboration among data scientists, ML engineers, and operations teams, and improving the reproducibility and security of deployed models \cite{ref-61}.

Within this landscape, Machine Learning Operations (MLOps) has emerged as a discipline that extends DevOps principles to the full lifecycle of ML models \cite{10081336}. MLOps aims to standardize, automate, and scale the development, validation, deployment, and monitoring of models in production. This includes practices such as Continuous Integration and Delivery (CI/CD), experiment traceability, environment management, and early error detection in deployed models. Shift Left becomes a foundational pillar for ensuring the security and reliability of production ML systems, with early-stage security practices such as data validation and vulnerability detection strengthening ML infrastructure and reducing the risk of failures in operational settings \cite{Huang2024}.

Despite growing interest in MLOps, practitioners face uncertainty regarding tool selection, integration strategies, and the maturity of available solutions. The current ecosystem offers a wide range of frameworks, from lightweight options like MLflow\footnote{https://mlflow.org} and Metaflow\footnote{https://metaflow.org} to more complex, production-oriented platforms such as Apache Airflow\footnote{https://airflow.apache.org/}, Amazon Sagemaker\footnote{https://aws.amazon.com/es/sagemaker/}, TFX\footnote{https://www.tensorflow.org/tfx}, and Kubeflow Pipelines\footnote{https://www.kubeflow.org/docs/components/pipelines/}. This diversity complicates decision-making for teams seeking to implement robust MLOps pipelines.

In order to provide guidance to practitioners in the field of MLOps, we present a practical and comparative evaluation of four representative tools: MLflow, Metaflow, Apache Airflow, and Kubeflow Pipelines. We implement pipelines using each tool, and define objective criteria to compare their performance, usability, flexibility, and adaptability. Our goal is to provide a clear, evidence-based perspective on the strengths and limitations of each solution.

Given the extant diversity of MLOps frameworks and technics, current literature has produced plenty reviews to tackle this complexity. However, most of them assess groups of academic works (In the range of a few tens) \cite{ZAROUR2025107733,10.1145/3747346,wazir2023mlopsreview,10011505,10855428,lima2022mlops,9582569,10.1145/3625289} and derive insights and recommendations. Even more, the abundance of this type of works has led to authors like Hanchuk and Semerikov \cite{hanchuk2025automating} to conduct a \textit{review of reviews}, in which analyze and provide a synthesis of the conclusions obtained by those reviews. Other approaches formalize the analysis of MLOps tools to increase their understandability by practitioners \cite{10440483, JOHN2025107725, 10081336}. Despite useful in a general level, these (meta)reviews fail to provide the focus we are aiming for in this work, namely to analyse specific tools, and  to provide precise, actionable and practical knowledge to the MLOps professionals. 

The remainder of this article is structured as follows: section \ref{sec:frameworks} describes the compared MLOps frameworks; section \ref{sec:experimental} provides the details of the experiments performed for the comparison, including the assessed features (Section \ref{sec:features}), use cases (Section \ref{sec:use-cases}), and deployment (Section \ref{sec:deployment}); section \ref{sec:results} provides the results of the experiments, and finally, section \ref{sec:conclusions} provides a wrapping conclusion and pointers for future directions.

\section{MLOps frameworks}\label{sec:frameworks}

In the development of this work, we have opted to assess open-source tools, which we belive are of most interest to the community, in contrast to commercial alternatives. These have been selected with the objective of covering all phases of the ML model lifecycle, following MLOps principles.

The tool selection was based on industry popularity, ease of integration, documentation, available support, and capacity to be deployed in production (scalable) environments. Additionally, we valued the possibility of automating workflows, versioning models and data, and performing testing and monitoring in a systematic manner. This section describes the main tools compared in this work. 

\subsection{MLflow}

MLflow is an open-source platform designed to manage the entire ML lifecycle. It was originally developed by Databricks, the company also behind Apache Spark, and is currently maintained by this company together with an active developer community. At the time of writing, the official GitHub repository has over 20,700 stars and more than 850 contributors, with frequent updates and a public roadmap. This combination of enterprise support and community contribution provides stability for the future.

It has become one of the most popular MLOps tools due to its flexibility and comprehensive set of features:

\begin{itemize}
    \item Tracking: MLflow's tracking component allows users to record and query experiments, including code, data, configuration and results. This facilitates tracking of model development progress, comparison of different experiments and ensures reproducibility. The MLflow Tracking component allows storing metrics, hyperparameters, auxiliary files and outputs from each execution, either in a local file system or in remote storage services such as Amazon S3, Google Cloud Storage or Azure Blob Storage. 
    \item Projects: MLflow organizes ML code into reusable and reproducible projects. Each project contains a self-contained Conda environment and a set of parameters, which simplifies the process of sharing and reproducing experiments across different environments.
    \item Models: MLflow provides a standardized format for packaging and versioning ML models. This allows models to be deployed across different platforms and runtime environments with minimal changes.
    \item Model Registry: MLflow includes a central model store that provides versioning, stage transitions, annotations and management of ML models. This allows controlling model stages (Staging, Production, etc.) facilitating the complete deployment lifecycle. This centralized management facilitates reproducibility, auditing and traceability of work.
\end{itemize}

MLflow offers a modular architecture that allows flexible management of both data and models and parameters associated with experiments. Although it can be used freely locally or on dedicated servers, Databricks offers native MLflow integration within its cloud platform, allowing companies to delegate infrastructure management, scalability, storage and security to this provider. This commercial version offers advantages such as multi-user collaboration, centralized access control and large-scale monitoring, oriented towards enterprise teams. This duality between open-source versions and managed commercial solutions has become a common strategy in the MLOps ecosystem, as it allows balancing the economic sustainability of the project with the principles of openness and collaboration inherent to free software.

Due to its modular approach and its capacity for integration with multiple tools in the ML ecosystem, MLflow has been adopted by companies such as Microsoft, Meta, Toyota or Accenture.

\subsection{Metaflow}

Metaflow is an open-source framework initially developed by Netflix, designed to simplify the creation and management of workflows in data science and ML. Its design focuses on ease of use for data scientists, allowing writing pipelines in Python naturally and scalably, without sacrificing advanced functionalities such as versioning, distributed execution or artifact tracking. 

Metaflow is licensed under Apache 2.0, and although Netflix continues to maintain it as an open source project, it is also supported by Outerbounds, a company founded by former Netflix employees. Outerbounds\footnote{Outerbounds: http://outerbounds.com} offers a commercial platform based on Metaflow that provides additional functionalities such as advanced monitoring, centralized control panel, enterprise support and managed scalability. This combination of community and commercial support has contributed to its maturity and adoption by other companies such as Warner Bros Discovery, Intel or Zillow.

Its main features include:

\begin{itemize}
\item Workflow definition: Workflows in Metaflow are directed graph of steps written to describe a ML pipeline. They defined through Python classes, where each pipeline stage is marked with the \texttt{@step} decorator. This approach allows structuring code in a readable and modular way.
\item Data and artifact management: Metaflow automatically manages data between pipeline steps. Variables created in one step can be used directly in the next without manual serialization, thanks to an internal automatic persistence system managed by Metaflow, which allows recovering artifacts between steps transparently through the client API (Metaflow Client). This favors traceability and reproducibility of workflows, as all artifacts are registered.
\item Local or distributed execution: Workflows can be executed locally or automatically scale on cloud services such as AWS Batch, Step Functions or Kubernetes, although their integration is especially optimized for AWS. This hybrid behavior allows developing locally and deploying in production without changing the code.
\item Versioning and tracking: Each execution generates a unique version that stores inputs, outputs, intermediate state and metadata, which facilitates debugging, comparison and reproduction of experiments.
\item Graph visualization: Metaflow provides a Web interface and tools to visualize the execution graph, view logs and navigate between previous executions.
\end{itemize}

The project is active and in constant evolution. At the time of writing, its official GitHub repository has over 8,900 stars, more than 110 contributors and frequent releases, with documentation and numerous examples. There is a growing community on Slack and specialized forums where users share best practices and solutions to common problems.

\subsection{Apache Airflow}

Apache Airflow is an open-source workflow orchestrator widely used in data engineering and data science environments. It was originally created at AirBnB in 2014 and subsequently donated to the Apache Software Foundation, where it is currently maintained as a top-level project under the Apache 2.0 license. Its main strength lies in the declarative definition of complex workflows through DAGs (Directed Acyclic Graphs), written in Python, which gives the user complete control over dependencies, execution, and task scheduling. Its main features include:

\begin{itemize}
\item DAG definition: Workflows are defined as DAGs in Python files. Each node represents a task, and the links mark dependencies. This allows designing complex processes in a modular and reusable way.
\item Scheduling and execution: Airflow allows executing tasks periodically (through Cron expressions, a syntax used to schedule recurring tasks at defined intervals) or manually. It uses a scheduler and a queue system to distribute tasks across multiple Workers, which facilitates parallelization and horizontal scaling.
\item State and data management: Airflow does not directly move data between tasks, but it does manage their state and dependencies through a metadata system stored in a relational database (Typically PostgreSQL or MySQL). Each task saves information such as its state (Success, failure, retry), duration, logs, etc.
\item Retries, SLA and alerts: Airflow offers advanced error control mechanisms like automatic retries, Service Level Agreement (SLA) management, email alerts or integration with external services like Slack or PagerDuty.
\item Visual interface: Its Web panel allows visualizing and controlling DAG execution, querying logs, retrying tasks and modifying executions in real time.
\item Extensibility and ecosystem: Airflow has a modular architecture based on plugins, custom operators and sensors. Additionally, it has hundreds of pre-built integrations with AWS, GCP, Azure, Spark, Hadoop services, among others.
\end{itemize}

\subsection{Kubeflow Pipelines}

Kubeflow Pipelines (KFP) is one of the main components of the Kubeflow project, an initiative originally driven by Google and now maintained by the open-source community together with companies like Arrikto or Canonical. Its objective is to facilitate the creation, execution and monitoring of ML pipelines in environments managed by Kubernetes. It is especially oriented towards production deployments, in the cloud or in local infrastructures with high scalability and control demands. 

Kubeflow is an open-source project with Apache 2.0 license. Although it is managed as a community initiative, it has been historically driven by major technology players such as Google, IBM or Red Hat. At the time of writing this work, the Kubeflow Pipelines repository has over 3,900 stars on GitHub and more than 450 contributors. At the same time, it maintains stable activity, with frequent releases, extensive documentation and multiple tutorials. The community is organized through the official Slack channel.

At the enterprise level, there are commercial platforms that offer Kubeflow as a managed service. Some notable examples are AWS SageMaker Kubeflow Components, Arrikto Kubeflow Enterprise or Canonical Charmed Kubeflow. These solutions allow deploying Kubeflow in a few steps, adding advanced functionalities such as corporate security, centralized monitoring and technical support.

Its main features include:

\begin{itemize}
\item Pipeline definition: Workflows can be defined imperatively through the Python SDK, or declaratively with YAML. Each pipeline is composed of components that encapsulate individual steps, allowing reuse and modular composition.
\item Execution on Kubernetes: Each pipeline component is executed as an independent container within a Kubernetes cluster. This allows isolation, scalability and replicability, natively leveraging cluster resources.
\item Data and artifact management: Kubeflow Pipelines uses an internal storage system (Based on MinIO or persistent volumes) to store inputs, outputs, artifacts, metrics and logs from each component. Additionally, it allows visualizing these outputs from the UI, which facilitates post-execution analysis.
\item Versioning and traceability: Each pipeline, component and experiment is registered with its version, parameters, execution time and results. This facilitates both execution comparison and scientific reproducibility or quality control in production.
\item Graphical interface: Kubeflow Pipelines includes a Web UI that allows executing workflows, viewing past executions, graphically inspecting the DAG, and querying metrics or artifacts.
\item Integration with other tools: Kubeflow Pipelines integrates natively with TensorBoard for metrics visualization and with Katib for automatic hyperparameter optimization. It can also connect with storage systems, monitoring tools like Prometheus or deployment tools like KServe.
\end{itemize}

\section{Experimental setup}\label{sec:experimental}

This section details the methodology used to evaluate the selected MLOps tools. The evaluation is structured around three main aspects: the key characteristics that define the usability and effectiveness of each tool, the practical use cases designed to test their functionality in real scenarios, and the deployment strategy for each tool.

\subsection{Assessed features}\label{sec:features}

To ensure an objective comparison of the selected MLOps tools (MLflow, Metaflow, Apache Airflow, and Kubeflow Pipelines), evaluation dimensions have been defined, 
where the maximum score for each criterion is 10 and the minimum is 0 (Based on the criteria defined on page 25 of \cite{ref-1}.). They are described as follows, including their relative weight (as a percentage):

\begin{itemize}
  \item Installation ease (15\%): This criterion considers the initial complexity to set up the tool. Simple installation favors rapid adoption by teams, reduces technical barriers, and accelerates the testing phase. 

  \item Configuration flexibility (15\%): Evaluates how adaptable the tool is to different scenarios, models, and workflows. Customization is key to adjusting to specific needs because  data and model pipelines vary greatly between projects. 

  \item Interoperability and integration with other tools (20\%): The ability to integrate with other tools is crucial, especially from a data engineering perspective. MLOps should easily integrate with orchestrators like Airflow, CI/CD tools like GitHub Actions, data storage services, and enterprise APIs. 

  \item Instrumentation complexity (15\%): The ease of connecting existing code with the tool, using clean APIs, intuitive syntax, and support for popular libraries (such as PyTorch, Keras, or scikit-learn).

  \item Result interpretability (15\%): The tool's ability to show clear dashboards, execution traceability, versioning, and model portability are fundamental for validating and reproducing results.

  \item Documentation and support (20\%): Clear documentation and an active community that allow to quickly resolve technical blocks and make the most of the tool's capabilities.
\end{itemize}

\subsection{Use cases}\label{sec:use-cases}

With the objective of evaluating the selected MLOps tools with real-world ML model lifecycle situations, two practical use cases were defined. Both cases were implemented and executed on each of the evaluated tools to observe differences in the assessed features, to compare how each tool responds to different technical challenges: custom vs. pre-trained models, structured data vs. free text, local vs. orchestrated pipelines, and metric storage vs. model versioning. Table \ref{tab:comparativa_casos_uso} shows the differences between the presented use cases.

\begin{table}
    \centering
\begin{tabular}{lll}
\textbf{Aspect}                                                         & \textbf{Digit classifier (MNIST)}                                                             & \textbf{Sentiment classifier (IMDB)}                                                              \\ \hline
\textbf{Data type}                                                      & Grayscale images (28x28 pixels)                                                               & Text (movie reviews)                                                                              \\ \hline
\textbf{Model used}                                                     & \begin{tabular}[c]{@{}l@{}}SimpleNN, DeepNN, CNN neural\\ networks in PyTorch\end{tabular}    & \begin{tabular}[c]{@{}l@{}}distilBERT (pre-trained Transformer\\ model)\end{tabular}              \\ \hline
\textbf{\begin{tabular}[c]{@{}l@{}}Primary\\ objective\end{tabular}}    & Handwritten digit classification                                                              & Sentiment analysis (positive/negative)                                                            \\ \hline
\textbf{\begin{tabular}[c]{@{}l@{}}Model\\ complexity\end{tabular}}     & \begin{tabular}[c]{@{}l@{}}Moderate (simple and complex\\ neural network models)\end{tabular} & \begin{tabular}[c]{@{}l@{}}High (pre-trained BERT model with\\ long text processing)\end{tabular} \\ \hline
\textbf{Task type}                                                      & Image classification                                                                          & Text classification                                                                               \\ \hline
\textbf{Data size}                                                      & \begin{tabular}[c]{@{}l@{}}MNIST dataset\\ (60,000 training images)\end{tabular}              & \begin{tabular}[c]{@{}l@{}}IMDB dataset\\ (500 stratified samples)\end{tabular}                   \\ \hline
\textbf{\begin{tabular}[c]{@{}l@{}}Pipeline\\ scalability\end{tabular}} & \begin{tabular}[c]{@{}l@{}}Low resource demands\\ (image processing)\end{tabular}             & \begin{tabular}[c]{@{}l@{}}High resource demands (text processing\\ with BERT)\end{tabular}       \\ \hline
\end{tabular}
    \caption{Use case comparison.}
    \label{tab:comparativa_casos_uso}
\end{table}

\subsubsection{Digit classifier with MNIST} 

The aim of this use case is to build an image classification model using the MNIST dataset\footnote{\url{https://huggingface.co/datasets/ylecun/mnist}}, which contains grayscale images of handwritten digits (0–9). Three different architectures have been implemented in PyTorch to observe how each tool manages experimentation and reproduction, covering a range of increasing complexity: 

\begin{itemize}
    \item SimpleNN (Simple neural network)\footnote{\url{https://labur.eus/szedzhwv}}: This is the most basic model. It uses two layers: one to process the image and another one to predict the digit. Each image is converted into a vector of 784 numbers and then passed to a function that activates neurons and allows learning patterns.

    \item DeepNN (Deep dense network)\footnote{\url{https://labur.eus/zdq2lbps}}: This model is a more advanced version, with more intermediate layers. Each layer progressively transforms the image into a more useful representation for predicting the correct digit. Additionally, a technique called Dropout is included that helps the model to avoid memorizing data and therefore learn more generally.

    \item CNN\_Model (Convolutional network)\footnote{\url{https://labur.eus/v8gnatra}}: This model is specifically designed to work with images. It uses convolutional layers that are capable of detecting visual patterns like edges or shapes. Techniques such as Batch Normalization are also used to stabilize training, and MaxPooling to reduce image size without losing important information.

\end{itemize}

\subsubsection{Sentiment classifier with IMDB and BERT} 

This use case applies Natural Language Processing (NLP) techniques for sentiment classification in movie reviews, using the IMDB dataset\footnote{\url{https://huggingface.co/datasets/stanfordnlp/imdb}}. Unlike the first use case, this use case introduces long texts, varied vocabularies, and complex linguistic structures, adding new challenges in terms of data processing.

To address this task, the pre-trained model \textit{distilBERT} has been employed, a reduced and optimized version of the Bidirectional Encoder Representations from Transformers (BERT) \cite{ref-133}. BERT was proposed by Google in 2018 as a model based on the Transformer architecture and revolutionized the NLP field by introducing the concept of bidirectional pre-training of text representations, allowing the model to understand word context considering content both before and after the word.

BERT is a computationally demanding model, with around 110 million parameters, making it expensive to run in resource-limited environments. For this, in this work, we have implemented distilBERT\footnote{\url{https://labur.eus/xzpt0axt}}, a lighter version that retains 97\% of BERT's performance on standard tasks, but is 40\% smaller and 60\% faster in inference. From an architectural perspective, distilBERT is based on the same principles as BERT, but eliminates some layers and redundant mechanisms to lighten the model.

This use case addresses text processing with distilBERT, fine-tuned for sentiment analysis, which introduces greater complexity both in terms of resources and pipeline configuration:

\begin{itemize}
\item Dependency and resource management: It requires downloading both the pre-trained model and its tokenizer, which increases loading time and memory usage, especially in resource-limited environments.
\item Specific preprocessing: Text cannot be processed with simple word tokenization. The model's tokenizer must be used, which transforms sentences into sequences of encoded subwords.
\item Inference pipeline configuration: It is necessary to create and configure a specialized sentiment analysis pipeline using the Transformers library. This involves managing parameters such as maximum length, truncation, and execution device (CPU or GPU).
\item Higher computational cost: Although distilBERT is more efficient than BERT, it remains significantly slower than traditional models like simple neural networks and it requires more resources for inference and training, which can affect pipeline scalability.
\end{itemize}

This use case has been key to comparing how different MLOps tools behave with more demanding models, analyzing their capacity to handle external dependencies, manage computational resources, and reproduce complex executions without loss of traceability or flexibility.

\subsection{Deployment}\label{sec:deployment}

The deployment process was designed with the objective of Continuously Integrating (CI) model training and evaluation, as well as their subsequent storage and tracking through MLOps tools. This process is essential to ensure that models and pipelines are easily reproducible, scalable, and manageable in production environments.

The complete source code is located in a GitHub repository\footnote{\url{https://github.com/Jonmaa/MLOps}}. The repository is structured according to the evaluated tools, including separate scripts for MLflow, Metaflow, Apache Airflow, and Kubeflow Pipelines. This organization allows for the execution of each pipeline separately and the clear observation of differences in integration and deployment. The use of a centralized repository ensures that all code versions are managed in a controlled manner, allowing complete traceability of changes, thus enabling identification of failures in case any error arises when making changes.

\begin{figure}[H]
    \centering
    \includegraphics[width=0.7\linewidth]{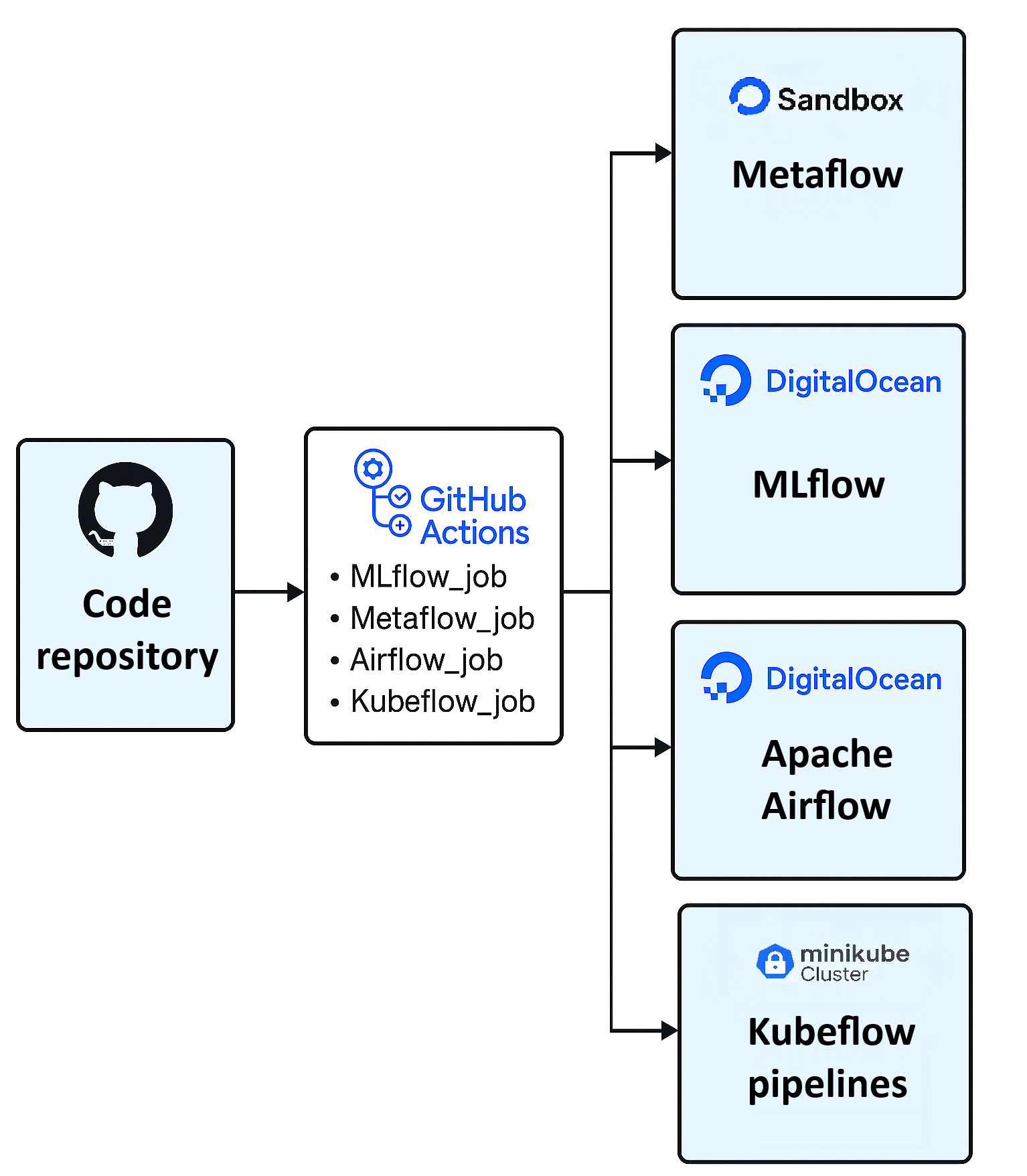}
    \caption{MLOps deployment architecture illustrating the evaluation workflow. Code changes in the repository trigger GitHub Actions jobs (MLflow\_job, Metaflow\_job, Airflow\_job, Kubeflow\_job) that deploy to respective environments: Metaflow Sandbox for prototyping, MLflow and Apache Airflow servers on DigitalOcean droplets, and Kubeflow Pipelines on a local Minikube cluster.}
    \label{fig:diagrama_resumen}
\end{figure}

Figure \ref{fig:diagrama_resumen} summarizes the process. The process begins when a code change is uploaded to the GitHub repository. An automated workflow was implemented with GitHub Actions\footnote{\url{https://labur.eus/mzhoohc0}} (\texttt{main.yml}), which is activated when detecting changes in model training and test files (\texttt{mlflowMain.py}, \texttt{mlflowTest.py}, \texttt{metaflowMain.py}, etc.). This workflow performs the following steps:

\vspace{0,5cm}
\begin{enumerate}
  \item Detect in which files the change has been made; depending on the file, one set of tasks or another will be executed (e.g. if a change has been made to an MLflow file, only the task corresponding to MLflow would be executed).
  \item if \textit{MLflow\_job}: Execute the files and upload artifacts to the server.
  \item if \textit{Metaflow\_job}: Execute the files and upload artifacts to Github Actions.
  \item if \textit{Airflow\_job}: Upload files to the server.
  \item if \textit{Kubeflow\_job}: Display message of change made to files. 
\end{enumerate}

This system ensures execution reproducibility and facilitates version control of both code and generated models, all from the same repository.

On regards to resources, two virtual machines were created in DigitalOcean (Droplets), on which both the MLflow and Airflow servers were installed. This way, the complete process can be performed: from making a change to the document at the request of another team member to having it uploaded directly to the server where the results can be viewed.

Metaflow and Kubeflow Pipelines were used in a different way due to their nature. Metaflow was assessed using the sandbox provided in the Cloud due to the impossibility to run a fully fledged installation in a DigitalOcean Droplet. KFP was deployed in a local installation, as it was run on top of a \textit{minikube} installation, which provided a local Kubernetes cluster.

Further detail about the deployment of these tools is provided in Appendix \ref{secA1}.

\section{Results}\label{sec:results}

This section presents the results obtained after the practical evaluation of the four selected MLOps tools: MLflow, Metaflow, Apache Airflow, and Kubeflow Pipelines. The assessment of each feature was based on the experience during the implementation of the use cases (MNIST and IMDB). The structure and weighting defined in section \textbf{\ref{sec:features}} were followed.

\subsection{Ease of Installation}

The initial installation of an MLOps tool can be the deciding factor between its adoption or rejection. Although this process is only performed once, it can consume significant resources if it requires complex configurations or advanced technical knowledge. This section evaluates both the time required and the clarity of the steps involved, the number of dependencies, and the ease of deploying the tool in local or cloud environments.

\textbf{MLflow} has proven to be one of the simplest tools to install, requiring the execution of a Python \textit{pip} command. After installation, a tracking server (a functional environment where results will be sent) can be configured by executing a single command, without requiring any additional complex configurations. \textbf{Score: 9}.

\textbf{Apache Airflow} requires a more complex installation, including setting up a database (SQLite or PostgreSQL), a virtual environment, and multiple components (Scheduler, Webserver, Worker). For this project, which aimed to develop a basic testing environment, the default database provided with the installation was used, following its official installation guide.Once everything is installed, the Airflow server can be launched with a single command. \textbf{Score: 8}.

\textbf{Metaflow} has a simple installation process for the most basic usage, but its user interface lacks official support and requires complex backend configurations (Metaflow service, UI, database, and storage). Installing Metaflow is as simple as running a Python \textit{pip} command. In this work, we attempted to deploy the UI locally following instructions from Netflix's repositories, but without success. Therefore, we opted to use the sandbox environment provided by them. \textbf{Score: 4}.

\textbf{Kubeflow Pipelines} was the most demanding in terms of installation. It requires deploying a Kubernetes cluster, configuring persistent storage, authentication, and multiple components (Argo Workflows, Katib, TensorBoard, etc.). This complexity represents a significant barrier to entry, especially for users without prior Kubernetes experience. The complexity of installing KFP can deter many companies from choosing this option unless they already have a cluster in place. \textbf{Score: 6}.

\subsection{Configuration flexibility}

Configuration flexibility refers to each tool’s ability to adapt to diverse workflows, parameterize experiments, reuse configurations, and define complex or conditional tasks. Both the expressive power of the system and its ease of practical use are evaluated.

\textbf{MLflow} provides a modular architecture built on two main components: the \textit{tracking server}, which stores metadata and artifacts from each run, and the \textit{Model Registry}, which manages versions of trained models.

Listing \ref{lst:establecer_experimento_mlflow} shows a code snippet in which, firstly, the URI of the server where results will be stored must be set, along with the name of the experiment. Then, experiments and logging parameters and metrics are defined with explicit calls. With this, experiments and runs are then organized in the Web interface of the server, accessible via a browser.

\begin{lstlisting}[caption={Connection with the tracking server for the MNIST model}, label = {lst:establecer_experimento_mlflow}]
mlflow.set_tracking_uri("http://11.22.33.44:5000")
mlflow.set_experiment("MNIST-Classification")
with mlflow.start_run() as run:
    mlflow.log_param("learning_rate", lr)
    ...
    for epoch in range(num_epochs):
        ...
        mlflow.log_metric("training_loss", avg_loss, step=epoch)
\end{lstlisting}

In a situation where, from any other script or environment, the latest approved version of the model can be retrieved using the line shown in Code Snippet {\ref{lst:mlflow_recover_model}. This model can then be integrated into inference, validation, or CI/CD workflows.

\begin{lstlisting}[caption={Loading the MNIST model from the Model Registry}, label = {lst:mlflow_recover_model}]
model = mlflow.pyfunc.load_model("models:/...")
\end{lstlisting}

Thanks to this architecture, MLflow enables experiment reproducibility, run comparison, version history tracking, and simplifies the model lifecycle in collaborative environments. However, MLflow does not provide a system for defining structured pipelines or conditional tasks. This kind of logic must be implemented externally using scripts or complementary tools. \textbf{Score: 8}.

\vspace{0.5em}
\textbf{Apache Airflow} allows defining complex workflows through DAGs, where each task represents an isolated unit that can depend on others. Model definitions are made using independent functions. In Airflow, tasks are encapsulated as Python functions and organized as operators, as shown in Code Snippet \ref{lst:airflow_operadores_tareas}.

\begin{lstlisting}[caption={Definition of MNIST model DAG with chained tasks}, label = {lst:airflow_operadores_tareas}]
with DAG(
    dag_id='mnist_classification_airflow',
    default_args=default_args,
    description='Train_and_evaluate_MNIST_with_PyTorch_(DeepNN)',
    schedule_interval=None,
    catchup=False,
    tags=['mnist','pytorch'],
) as dag:

    t1 = PythonOperator(
        task_id='train_model',
        python_callable=train_model,
        provide_context=True,
    )

    t2 = PythonOperator(
        task_id='evaluate_model',
        python_callable=evaluate_model,
        provide_context=True,
    )

    t1 >> t2
\end{lstlisting}

In the context of Apache Airflow, \texttt{ti} variables refer to instances of the \texttt{TaskInstance} class. This class represents a specific execution of a DAG task and is commonly used for communication between tasks — this is known as \textit{XCom} (Cross-communication). Airflow stores XCom values in its internal database, ensuring traceability and persistence throughout the DAG’s execution.

This approach grants a high degree of control over each stage, allows for the use of multiple environments (Docker, Bash, Python), and scales to complex architectures. \textbf{Score: 9}.

\vspace{0.5em}

\textbf{Metaflow} provides a clear and flexible design for defining workflows. Each step is defined as a method decorated with \texttt{@step} within a \texttt{FlowSpec} class. Code Snippet \ref{lst:metaflow_estructura}, shows the typical structure.

\begin{lstlisting}[caption={Training pipeline for the MNIST model using Metaflow}, label = {lst:metaflow_estructura}]
class MNISTFlow(FlowSpec):

    @step
    def start(self):
        self.next(self.load_data)

    @step
    def load_data(self):
        ...
        self.next(self.train_model)

    @step
    def train_model(self):
        self.model = DeepNN()
        ...
        self.next(self.evaluate_model)
\end{lstlisting}

This structure promotes separation between data logic and flow control, keeping the code readable and easy to debug. Metaflow allows defining hyperparameters and configurations as class attributes using the \texttt{Parameter} class, as shown in Code Snippet \ref{lst:metaflow_parameter}. This facilitates reuse and allows executing runs with different values without modifying the source code.

\begin{lstlisting}[caption={Configurable parameters, example from MNIST model}, label = {lst:metaflow_parameter}]
batch_size = Parameter('b_size', default=64, help='Batch size')
lr = Parameter('l_rate', default=0.001, help='Learning rate')
epochs = Parameter('epochs', default=10, help='Epoch number')
\end{lstlisting}

During execution, Metaflow automatically stores the intermediate state of each step. It also allows generating visual reports using the \texttt{Cards} system, which can include numerical results or embedded graphs. These cards are automatically stored in Metaflow’s backend and can be accessed via the web interface or CLI, facilitating interpretation of results and tracking of past experiments.

Each flow run generates a unique identifier and can later be retrieved to reanalyze results, access intermediate data, or repeat specific steps. This provides a robust solution for scientific traceability. Overall, Metaflow combines ease of use with powerful tracking and modularity capabilities, making it especially useful in exploratory and rapid prototyping contexts. \textbf{Score: 9}.

\vspace{0.5em}

\textbf{Kubeflow Pipelines} allows defining workflows through YAML or a Python DSL, where each component is defined as a function. As an example, Code Snippet \ref{lst:kfp_estructura} shows the sentiment analysis workflow, structured with two tasks.

\begin{lstlisting}[caption={General structure of the MNIST model pipeline in KFP}, label = {lst:kfp_estructura}]
apiVersion: argoproj.io/v1alpha1
kind: Workflow
metadata:
  generateName: mnist-classification-
spec:
  entrypoint: mnist-pipeline
  templates:
  - name: mnist-pipeline
    dag:
      tasks:
      - name: train-model
        template: train-model
      - name: evaluate-model
        template: evaluate-model
        dependencies: [train-model]
\end{lstlisting}

Each step uses a Docker image (in this case, \texttt{pytorch/pytorch}) and executes Python scripts generated dynamically at runtime.

Once deployed, the pipeline can be executed from the Kubeflow graphical interface, where steps, generated artifacts, logs, and metrics from each run are displayed. This facilitates monitoring, debugging, and comparing model versions in production environments. While this setup offers great control, it requires prior experience with containers, volumes, and Kubernetes. \textbf{Score: 9}.

\subsection{Interoperability and integration with other tools}

In a modern MLOps system, no tool operates in isolation. Interoperability is essential to guarantee scalability, automation, and operational efficiency. This section evaluates each tool's level of integration with CI/CD systems such as GitHub Actions, storage services, databases, and cloud platforms.

\textbf{MLflow} provides a well-documented REST API and a CLI that enables programmatic registration of artifacts, models, and metrics. In this work, the Tracking Server was successfully integrated with GitHub Actions, using \texttt{rsync} and SSH keys to synchronize artifacts generated in the execution environment with the remote server hosting the Tracking Server. In addition, using the Python API, the model was automatically transitioned to the \texttt{Production} stage in the Model Registry (\texttt{mlflow.register\_model} and \texttt{MlflowClient.transition\_model\_version\_stage}). However, MLflow does not offer native connectors for enterprise databases or messaging systems, so integration with external services must be developed programmatically. \textbf{Score: 7}.

\textbf{Apache Airflow} stands out for its strong integration capabilities. As a widely adopted workflow orchestrator in enterprise settings, it includes a large number of native operators to interact with services such as PostgreSQL, MySQL, Amazon S3, Docker, Spark, among others. In this work, Airflow orchestrated the training and evaluation of the MNIST and IMDB models using \texttt{PythonOperator} and the XCom system to pass artifacts between tasks. DAG synchronization with GitHub Actions was achieved without difficulty, demonstrating good compatibility with CI/CD systems. Thanks to its modular architecture and extension support, Airflow is especially suitable for heterogeneous enterprise workflows. \textbf{Score: 9}.

\textbf{Metaflow} is designed to facilitate rapid experimentation, particularly in the AWS ecosystem. It offers direct integration with Amazon S3 and Step Functions, and can execute on AWS Batch or locally. In this work, it was used locally, but integration with GitHub Actions was validated to automate runs. Inter-step data transfer is handled automatically. Although it does not provide a system of operators or native connectors like Airflow, it allows straightforward custom integrations in Python. Its compatibility with external APIs and storage systems depends on the flow code itself, which provides flexibility at the cost of additional developer effort. \textbf{Score: 7}.

\textbf{Kubeflow Pipelines} is specifically designed for Kubernetes-based environments, which determines its integration approach. Each pipeline component is defined as a Docker container and can communicate with others via shared volumes. It also supports configuration of \texttt{artifacts}, \texttt{secrets}, and services such as Katib (for hyperparameter optimization), TensorBoard, or MinIO. Its integration with cloud platforms such as GCP, AWS, or Azure is natural, since it shares infrastructure with cloud-native environments. However, this power comes with a steep learning curve, particularly around cluster configuration and YAML manifests. \textbf{Score: 8}.

\subsection{Code instrumentation complexity}

This section analyzes the difficulty of integrating each tool with project source code. We consider API readability, the number of lines required to instrument models, compatibility with common libraries (PyTorch, scikit-learn, TensorFlow), and the degree of intrusiveness in the original workflow.

\textbf{MLflow}: It features a simple, direct API for logging parameters, metrics, and models. Integration with Python code is minimal and does not require significant changes to model logic. In Listing \ref{lst:mlflow_register_params}, each run is explicitly registered with \texttt{start\_run()}, and training is monitored with only a few lines.

\begin{lstlisting}[caption={Registering parameters and metrics in MLflow for the IMDB model}, label = {lst:mlflow_register_params}]
with mlflow.start_run():
    mlflow.log_param("model_name", model_name)
    mlflow.log_param("max_length", 512)
\end{lstlisting}

The model is saved using the \texttt{PyFunc} wrapper, enabling reuse and deployment (see Listing \ref{lst:mlflow_imdb_pyfunc}).

\begin{lstlisting}[caption={Registering the model as a PyFunc for the IMDB model}, label = {lst:mlflow_imdb_pyfunc}]
 mlflow.pyfunc.log_model(
        artifact_path="sentiment_model",
        python_model=SentimentWrapper(model_name),
        input_example=example_df,
        signature=sig
    )
\end{lstlisting}

\noindent Compatibility with PyTorch, scikit-learn, and pandas is native and integrates seamlessly with the Python ecosystem. Instrumentation requires little additional code and imposes no particular structure. \textbf{Score: 8}.

\textbf{Apache Airflow}: Training logic must be encapsulated in independent functions that are called from the DAG as tasks using \texttt{PythonOperator}. This separation implies some duplication and forces decoupling of logic from control flow. For example, in Listing \ref{lst:airflow_train_model}, the function \texttt{train\_model} includes the full MNIST training logic. If another dataset were to be trained, the user would either implement a new, nearly identical function or further factor out the dataset-loading part into separate functions per dataset.

\begin{lstlisting}[caption={Training the MNIST model within a task}, label = {lst:airflow_train_model}]
def train_model(**kwargs):   
    transform = transforms.Compose([transforms.ToTensor()])
    train_ds = datasets.MNIST(root='/tmp/mnist_data', ...)
    train_loader = torch.utils.data.DataLoader(...)
    ...
    for epoch in range(10):
        for images, labels in train_loader:
            outputs = model(images)
            loss = criterion(outputs, labels)
            ...
    ti = kwargs['ti']
    ti.xcom_push(key='model_path', value=model_path)
\end{lstlisting}

In addition, \texttt{XComs} are needed to pass information between tasks, such as the model path. While this provides granularity and control, it also increases instrumentation overhead. \textbf{Score: 6}.

\textbf{Metaflow}: It offers a very natural instrumentation centered on the experiment's logical flow. Each step is defined with the \texttt{@step} decorator, and the entire pipeline is expressed as a class derived from \texttt{FlowSpec}. The model's internal structure does not need to be modified: data flow between steps is automatic so that variables defined in one step can be reused in later steps. Listing \ref{lst:metaflow_estructura_ampliada} provides a sample of this structure.

\begin{lstlisting}[caption={Modular structure of the IMDB flow}, label = {lst:metaflow_estructura_ampliada}]
class IMDBSentimentFlow(FlowSpec):

    @step
    def start(self):
        self.next(self.load_model)
        ...
 
    @step
    def load_model(self):
        self.tokenizer = AutoTokenizer.from_pretrained("distilbert-
        base-uncased-finetuned-sst-2-english")
        ...
        self.next(self.predict)
    
    @card
    @step
    def predict(self):
        texts = self.df_small.text.tolist()
        labels = self.df_small.label.tolist()
        preds, scores = [], []
        embeddings = []

        for txt in texts:
            ...
\end{lstlisting}

Metaflow is fully compatible with PyTorch, scikit-learn, and TensorFlow because it does not impose any specific ML library and allows working directly with custom classes and structures. In this project we used it with PyTorch without impediments, preserving the usual training structure. Training logic can remain the same as in a traditional script, minimizing intrusiveness and facilitating porting of existing models. \textbf{Score: 9}.

\textbf{Kubeflow Pipelines}: Instrumentation in KFP requires defining each component as a Docker container or as a DSL-decorated function. In this project, we opted for YAML manifests that run scripts dynamically generated with \texttt{cat}. This forces packaging of training logic into standalone files.

KFP is compatible with any library included in the corresponding Docker image (PyTorch, scikit-learn, TensorFlow, ...). However, the need to define artifacts, paths, volume mounts, and explicit inputs/outputs adds significant overhead. Compared to MLflow or Metaflow, the workflow is more intrusive: the model must be fully decoupled from the traditional Python environment and packaged for container execution. This architecture is very powerful for production, but less agile for development. \textbf{Score: 5}.

\subsection{Result interpretability} \label{sec:webUIs}

This section evaluates each tool's ability to clearly present experiment results, facilitate run traceability, compare iterations, and enable reuse and deployment of models across environments. These features are particularly relevant in collaborative, scientific validation, and production contexts. 

\textbf{MLflow} provides an intuitive graphical interface through its \texttt{Tracking UI}, where each experiment run can be explored: parameters used, recorded metrics, generated artifacts (such as models or images), and model versions. This hierarchical structure enables full traceability of the model lifecycle. Comparisons between runs are performed directly in the interface by selecting multiple runs.

MLflow includes a \texttt{Model Registry} where models can be versioned, documented, and promoted to stages such as \texttt{Staging} or \texttt{Production}. Being based on the \texttt{pyfunc} format, models can be exported to other formats such as ONNX or TorchScript, or served as a REST API using \texttt{mlflow serve}. \textbf{Score: 9}.

\textbf{Apache Airflow} provides a comprehensive control panel where one can visualize the execution history of DAGs, per-task execution times, dependencies, and logs for each step . The interface is useful for debugging and operational control, but it is not specialized for ML; it does not include native visualization of ML metrics, run comparison, or model management. 
Any interpretability elements (plots, reports, metrics) must be generated and saved manually within tasks—for example, exporting images or CSV files. \textbf{Score: 8}.

\textbf{Metaflow} has a simple and intuitive user interface, listing runs and allowing navigation into each to inspect failures or outputs. Like Apache Airflow and KFP, Metaflow can visualize the process as a DAG and it includes \texttt{Cards}, which allow generating HTML reports embedded in each flow step. These reports can contain rich text (Markdown), images, plots, and real-time metrics. 

Runs are fully traceable from the CLI or from a backend (such as AWS Step Functions), enabling reproduction of specific steps or comparison between runs. Although it lacks a unified Web panel like MLflow, its card system is flexible and well suited for prototyping and visual analysis. \textbf{Score: 7}.

\textbf{Kubeflow Pipelines} provides a visual interface to track pipeline execution step-by-step, inspect generated artifacts (e.g., models or metrics), consult per-task logs, and visualize the inputs/outputs defined in YAML. Tasks can also be visualized as a graph.

However, KFP does not provide direct integration with ML frameworks to log metrics or compare runs automatically. If results are to be visualized, scripts must explicitly export metrics in formats such as JSON or CSV, or visualizations that can then be rendered by additional tools (TensorBoard, Grafana, etc.). This lack of tight ML coupling implies more manual effort but also greater flexibility. \textbf{Score: 7}.

\subsection{Documentation and community support}

The quality of technical documentation and community support are determining factors for effective adoption of an MLOps tool. These aspects directly affect the initial learning curve, the ability to solve technical issues autonomously, and long-term project sustainability. Good documentation, accompanied by clear examples and an active community, can make a substantial difference in real development environments.

\textbf{MLflow}: It provides very comprehensive and well-structured official documentation, covering installation and basic use of the tracking server as well as model management in production, including integration with frameworks such as TensorFlow, PyTorch, scikit-learn, XGBoost, or Docker. It also includes a detailed API Reference and a guide for using the REST interface. Backed by Databricks, it benefits from periodic technical publications, an active GitHub community (over 20.4K stars), and tutorials on platforms such as Medium, YouTube, or Towards Data Science. \textbf{Score: 9}.

\textbf{Apache Airflow}: One of the most established tools in the DevOps ecosystem and usable in MLOps scenarios, although not as deeply ML-focused as other tools. Its official documentation is extensive, with guidance for installation, configuration, advanced DAG usage, connection to external services, and deployment in cloud or containers. The API Reference is well organized and details the available operators and hooks. The community is very active, with an official forum explaining its use with Astro, a tool that helps teams focus on data pipelines—build, run, and observe data in one place. There is also support through Stack Overflow, Slack channels, conferences, and training on platforms like Udemy or Coursera. Airflow is backed by the Apache Software Foundation, ensuring stability and ongoing maintenance. \textbf{Score: 9}.

\textbf{Metaflow}: The official documentation is clear, concise, and focused on practical framework usage. It includes a guide to core concepts, flow examples, AWS integration (S3, Batch, Step Functions), and documentation for visualization with \texttt{cards}. However, coverage is more limited than in other tools, especially for non-AWS environments. The community is small but active, with primary support in its GitHub repository, where issues are managed and updates published. Additional resources exist in blogs, YouTube, and Medium. Despite a good technical foundation, documentation could be more extensive and include UI installation and more advanced use cases. \textbf{Score: 7}.

\textbf{Kubeflow Pipelines}: Official documentation is broad and detailed, but also complex. Prior knowledge of Kubernetes, YAML, and containers is required to fully benefit. The site covers pipeline definition with the Python DSL, creation of custom components, and deployment on cloud clusters. However, resources are spread across versions and sometimes outdated. The Kubeflow community is sizable, with support on GitHub, Google forums, Stack Overflow, and Slack groups. Commercial initiatives such as Arrikto and Canonical also provide support and extended documentation. \textbf{Score: 7}.

\subsection{Summary of results}

Table \ref{tab:comparativa} summarizes the scores obtained by each MLOps tool evaluated in this project: MLflow, Metaflow, Apache Airflow, and Kubeflow Pipelines. The scores are based on hands-on experience during the development of the use cases (MNIST and IMDB), deployment in local and cloud environments, and technical analysis of each characteristic.

\begin{table}
\caption{Updated comparative evaluation matrix}
\label{tab:comparativa}
\begin{tabular}{|c|c|c|c|c|c|}
\hline
\textbf{Criterion} & \textbf{Weight} & \textbf{MLflow} & \textbf{Metaflow} & \textbf{Airflow} & \textbf{Kubeflow} \\
\hline
Ease of installation         & 15\%  & 9 & 4 & 8 & 6 \\ \hline
Configuration flexibility    & 15\%  & 8 & 9 & 9 & 9 \\ \hline
Interoperability             & 20\%  & 7 & 7 & 9 & 8 \\ \hline
Code instrumentation         & 15\%  & 8 & 9 & 6 & 5 \\ \hline
Interpretability             & 15\%  & 9 & 7 & 8 & 7 \\ \hline
Documentation / support      & 20\%  & 9 & 7 & 9 & 7 \\ \hline
\textbf{Total}               & 100\% & \textbf{8.30} & \textbf{7.15} & \textbf{8.25} & \textbf{7.05} \\
\hline
\end{tabular}
\end{table}

As shown, \textbf{MLflow} and \textbf{Apache Airflow} achieve the highest overall scores with a weighted average of 8.30. MLflow stands out for ease of use, strong documentation, and visual interpretability—an ideal option for agile, collaborative projects. Airflow excels in interoperability, flexibility, and community support, making it a robust tool for complex enterprise environments.

\textbf{Metaflow} provides an excellent developer experience and code instrumentation due to its modular, flow-oriented design. However, its score is affected by the difficulty of installing the UI and limited support beyond the AWS ecosystem.

\textbf{Kubeflow Pipelines} shows balanced performance across most criteria, especially flexibility and integration with Kubernetes. Nevertheless, its complex installation and instrumentation overhead decrease general usability, positioning it better for teams with prior cloud-native experience.

\subsection{Recommendations by usage context}

Beyond the numeric results, choosing an MLOps tool in real settings depends heavily on team profile, technical expertise, available infrastructure, and project objectives. The following recommendations are based on our experience during this work:

\begin{itemize}
    \item \textbf{MLflow} is well-suited for small teams or individual developers, especially in early project stages. Its easy installation, native integration with the Python ecosystem, and strong documentation make it ideal for experiment tracking, model versioning, and reproducibility without complex infrastructure. It is particularly useful in academic contexts or functional prototypes.

    \item \textbf{Apache Airflow} stands out for orchestration capabilities and the large number of native integrations. It is recommended in more mature environments where task dependencies, scheduled executions, or complex data flows must be controlled. While the learning curve is steeper, its maturity and industrial adoption make it a solid option for enterprises with DevOps or data engineering teams.

    \item \textbf{Metaflow} offers a balanced proposal between simplicity and power, with a clear syntax oriented to the experiment's logical flow. It is especially helpful in exploratory and model validation projects where traceability, agile prototyping, and modularity are valued. However, full installation—particularly the web interface—can be challenging without using the sandbox or AWS infrastructure, which may limit adoption in other contexts.

    \item \textbf{Kubeflow Pipelines} targets cloud production environments, especially those with an existing Kubernetes cluster. Its containerized, component-oriented architecture allows scaling complex flows, but installation, configuration, and usage require advanced technical knowledge. It is therefore more oriented to organizations with cloud-native experience and needs for high availability or task isolation.
\end{itemize}

In addition to these profile-based recommendations, the following cross-cutting aspects should be considered:

\begin{itemize}
\item \textbf{Learning curve:} MLflow and Metaflow offer a gentler onboarding for users experienced in Python and ML, allowing integration of experiments with minimal changes to base code. For example, in MLflow it suffices to wrap the training block in an experiment run with \texttt{mlflow.start\_run()} and log parameters/metrics with calls such as \texttt{mlflow.log\_param()} or \texttt{mlflow.log\_metric()}, as shown in Listing \ref{lst:mlflow_snippet}.

\begin{lstlisting}[caption={MNIST training snippet with MLflow}, label={lst:mlflow_snippet}]
with mlflow.start_run():
    mlflow.log_param("learning_rate", lr)
    for epoch in range(num_epochs):
    ...
        mlflow.log_metric("training_loss", avg_loss, step=epoch)
\end{lstlisting}

By contrast, Apache Airflow requires explicitly defining DAGs with multiple independent tasks and additional configurations, as shown in Listing \ref{lst:airflow_dag}.

\begin{lstlisting}[caption={Definition of an MNIST DAG in Airflow}, label={lst:airflow_dag}]
with DAG(dag_id="mnist_classification_airflow", ...) as dag:
    t1 = PythonOperator(
        task_id="train_model",
        python_callable=train_model,
        ...
    )
    t2 = PythonOperator(
        task_id="evaluate_model",
        python_callable=evaluate_model,
        ...
    )
    t1 >> t2
\end{lstlisting}

In KFP, even simple tasks must be encapsulated as containers defined in YAML files, which is an additional barrier without prior Kubernetes experience. Listing \ref{lst:kubeflow_yaml} shows a basic pipeline example.

\begin{lstlisting}[caption={Simplified IMDB pipeline example in Kubeflow}, label={lst:kubeflow_yaml}]
- name: sentiment-pipeline
    dag:
      tasks:
      - name: fetch-dataset
        template: fetch-dataset
      - name: analyze-with-model
        template: analyze-with-model
        dependencies: [fetch-dataset]
        arguments:
          artifacts:
          - name: imdb-dataset
            from: "{{tasks....}}"
\end{lstlisting}

These differences in initial complexity make MLflow and Metaflow more suitable for early phases or users without DevOps experience, while Airflow and Kubeflow are oriented to professional environments requiring greater control, scalability, or integration with other systems.

\item \textbf{Reproducibility and traceability:} Although all four tools allow recovery of prior runs and configuration logging, the level of integration, automation, and usability varies considerably. With MLflow, traceability is fully integrated. Each run is recorded within an experiment, including parameters, metrics, artifacts, and execution environment. These elements are organized with unique identifiers and can be queried via the web UI, CLI, or Python API. The entire tracking of a run is encapsulated in a \texttt{mlflow.start\_run()} block (see Listing \ref{lst:mlflow_snippet}), which simplifies comparison, repetition, and debugging of experiments. Metaflow also offers robust and automatic traceability. Each flow run stores parameters, outputs, and intermediate artifacts generated at each step, which can later be retrieved from any previous execution. The client interface (Metaflow Client API) allows navigation across versions and reuse of past outputs—particularly helpful in iterative development. 

By contrast, Apache Airflow and Kubeflow Pipelines primarily target task orchestration. Airflow automatically records task logs and status, and since version 3.0 it includes support for DAG versioning and tracking changes in their definition. Kubeflow provides traceability via ML Metadata (MLMD) to record parameters, metrics, artifacts, and relationships between runs. However, enabling this requires configuring components such as databases (PostgreSQL or MySQL), persistent storage, and access to volumes in Kubernetes—adding considerable complexity for teams without prior experience in cloud infrastructure or DevOps.

\item \textbf{Sustainability and support:} MLflow and Airflow have large communities and strong institutional backing, which ensures long-term maintenance. MLflow is developed by Databricks and follows a hybrid model: a very active open-source version and an enterprise version integrated into its commercial platform. The official documentation is extensive, with detailed examples and varied use cases. It also has native integrations with libraries such as PyTorch, scikit-learn, and Hugging Face, among others, which significantly reduces adoption effort. Airflow is a project of the Apache Software Foundation, with frequent updates and multiple corporate contributors. Its modular, extensible architecture has enabled adoption in large organizations. There are many community operators (for GCP, AWS, Docker, Slack, ...) and actively maintained plugins. 

In the case of Kubeflow, while backed by companies such as Google, Canonical, or Arrikto, its community is more fragmented and user experience varies across versions and environments. Initial configuration and deployment require advanced infrastructure knowledge, which can hinder adoption by teams with less Kubernetes or distributed-systems experience. Metaflow, originally developed by Netflix, has an active and growing community, especially in academic environments and startups. However, its ecosystem is still more limited, and some features (such as the Web dashboard or AWS-based deployment) are primarily aimed at managed cloud environments. Official documentation is good but not as extensive as MLflow or Airflow, and many solutions are found in repositories or unofficial forums.
\end{itemize}

These considerations complement the quantitative evaluation and can guide future work or technology adoption decisions depending on the specific usage context.

\section{Conclusions}\label{sec:conclusions}

This work has presented an empirical, comparative evaluation of four representative tools from the MLOps ecosystem: MLflow, Metaflow, Apache Airflow and Kubeflow Pipelines. Using two concrete use cases—image classification on MNIST and sentiment analysis on IMDB—we examined, via practical implementation, how these frameworks support the construction, execution and management of ML pipelines.

To ensure a rigorous comparison, we defined a set of objective evaluation criteria (installation effort, traceability, visualization, integration, documentation, etc.) and applied them consistently to each tool. The experimental work uncovered not only the main capabilities of each framework but also their strengths and weaknesses in realistic development contexts.

Our key findings are as follows: MLflow proved highly accessible for experiment tracking and model registration, distinguished by its low operational complexity and smooth integration with Python-based projects. Metaflow excelled at rapid prototyping of complex flows, offering a declarative and reproducible programming model. Apache Airflow demonstrated strong capabilities for task orchestration, albeit with a steeper learning curve and more operational overhead. Kubeflow Pipelines emerged as a robust option targeted at cloud and production deployments, but its installation and maintenance are more demanding in terms of infrastructure.

In terms of real practices, this study provides empirical guidance for engineers and researchers choosing MLOps tooling:

\begin{itemize}
  \item If rapid experiment tracking and low operational overhead are priorities, MLflow is a pragmatic choice for Python-centric projects.
  \item For agile prototyping of complex pipelines with readable code, Metaflow offers an effective developer experience.
  \item When robust, schedule-driven orchestration is required, Airflow provides expressive DAG semantics and mature operational features, at the cost of increased configuration and maintenance effort.
  \item For production-grade, Kubernetes-native pipelines with strong support for cloud-native integrations, Kubeflow Pipelines is appropriate, provided the organisation can invest in the necessary infrastructure and operational expertise.
\end{itemize}

In summary, this work provides a structured, practical, and comparative perspective on implementing MLOps using current tools, offering readers a solid foundation for approaching machine learning projects with scalable, reproducible, and maintainable methodologies.

Future work will focus on extending the current study beyond the core features evaluated in this comparison. Several advanced capabilities of the analyzed tools were not implemented due to time, resource, or setup constraints, but they represent valuable directions for deepening the analysis. Examples include enabling real-time model serving in MLflow through its REST API, configuring automated alerting mechanisms in Apache Airflow, and integrating hyperparameter optimization experiments in Kubeflow Pipelines via Katib. Implementing these features would allow future evaluations to emulate production-grade environments more closely, assessing aspects such as deployment readiness, operational monitoring, and automated model improvement.

A second line of future work involves scaling the experimental infrastructure to cloud-managed or distributed Kubernetes environments. Deploying pipelines on services such as Amazon EKS, Azure AKS, or Google GKE would provide insights into scalability, fault tolerance, and resource allocation under realistic workloads. Although managed services like AWS SageMaker were excluded in this study due to cost and vendor lock-in considerations, including them in future research would offer a more comprehensive perspective on the performance, interoperability, and operational costs of production-level MLOps systems.

Finally, the evaluation could be broadened by incorporating additional tools from the evolving MLOps ecosystem. Platforms such as Weights \& Biases, DVC, or MLReef could enrich the comparative landscape by highlighting differences in tracking capabilities, data versioning, collaboration support, and workflow integration. Expanding the toolset in this way would strengthen the generalizability of the results and help map how emerging open-source and commercial solutions position themselves across the MLOps lifecycle.

\section*{Data availability}
The supplementary material is available in the following GitHub repository: \url{https://github.com/Jonmaa/MLOps}.

\bmhead{Acknowledgements}

This work has been partially supported by the XWAVE research grant (KK2025/00056), funded by the Basque Government.












\begin{appendices}

\section{Details on tool deployment}\label{secA1}

This appendix extends the deployment descriptions provided in Section \ref{sec:deployment}. Code files mentioned in this section are available in the code repository that accompanies this manuscript.

\paragraph{MLflow}

To centralize experiment tracking and versioning, the MLflow server was deployed on a DigitalOcean droplet with 1 vCPU, 1 GB of memory, and 25 GB of SSD. On this server, MLflow was installed using Python 3.11 with the following configuration:

\begin{enumerate}
  \item Metadata backend: a local SQLite database (\texttt{mlflow.db}) is used as metadata store, created when executing the command to launch the server.
  \item Artifact store: artifacts are saved in the \texttt{/tmp/mlflow/artifacts} directory.
  \item HTTP exposure: the tracking server is launched with:
    \begin{verbatim}
     mlflow server \
        --backend-store-uri sqlite:////tmp/mlflow/mlflow.db \
        --default-artifact-root file:/tmp/mlflow/artifacts \
        --host 0.0.0.0 \
        --port 5000
    \end{verbatim}
    so that it remains accessible on port 5000 of the machine.
\end{enumerate}

The server is primarily responsible for:
\begin{itemize}
  \item Automatically register experiments, metrics, artifacts, and parameters from each execution.
  \item Persistently storing generated models.
  \item Serving as a centralized point for trained model management, facilitating their comparison and retrieval.
\end{itemize}

The droplet was manually configured with the necessary dependencies and connects directly with scripts executed by GitHub Actions, which use MLflow Tracking to register results remotely.

The MLflow job of the workflow defined in \texttt{.github/workflows/main.yml} is responsible for:

\begin{enumerate}
  \item Training and registration (\texttt{mlflowMain.py}): when a push is made to the \texttt{main} branch or a pull request is performed, the runner installs dependencies, trains the model, and registers parameters, metrics, and versions in MLflow's Model Registry.
  \item Secure synchronization: \texttt{rsync} is performed on artifacts from \texttt{/tmp/mlflow/artifacts} from the runner to the droplet, keeping the server's artifact store updated. \texttt{rsync} is executed securely through a SSH connection (Key stored in GitHub secrets). 
  \item Validation (\texttt{mlflowTest.py}): after artifact upload, the test script is executed: connecting to the same \texttt{MLFLOW\_TRACKING\_URI},
  evaluating the model against MNIST test, and registering final metrics.
\end{enumerate}


This setup allows a completely automated flow of training, registration, deployment, and validation, all orchestrated from GitHub Actions and hosted on an external server, facilitating collaborative access and experiment versioning with MLflow.

\paragraph{Airflow}

For task orchestration and MLOps pipeline automation, an Apache Airflow server was deployed on a DigitalOcean droplet with 2vCPUs, 4 GBs of memory, and 25 GBs of SSD. Airflow was installed in a virtual environment with Python 3.11, following the official documentation.

The basic configuration includes:

\begin{enumerate}
  \item Database: Airflow was configured with a SQLite database for the metastore, located at \texttt{/home/airflow/airflowServer/airflow.db}. It is generated automatically with the first server execution.
  \item DAGs folder: this is Airflow's entry point for loading Python scripts that define automated tasks.
  \item Service startup: Airflow is launched manually or as a persistent service through systemd supervision, and exposes the Web interface on port \texttt{8080}.
  
\end{enumerate}

On this server, the necessary DAGs are hosted to automate and control the experiment lifecycle. The key functionalities that Airflow allows for are:

\begin{itemize}
  \item Define scheduled workflows (DAGs) to perform tasks such as preprocessing, training, and evaluation.
  \item Control dependencies between tasks, execution times, and retries in case of failure.
\end{itemize}

The DAG code is hosted in the \texttt{/airflowServer/dags} directory, being able to then execute them from the user interface, allowing the user to observe how the different sections develop.

The Airflow job defined in \texttt{.github/workflows/main.yml} is responsible for:

\begin{enumerate}
  \item DAG upload: when a change is detected in the \texttt{Airflow/} folder of the repository, this job is activated. An SSH connection is established to the Airflow server and synchronization of all \texttt{.py} files is performed via \texttt{rsync}, copying them to the \texttt{/home/airflow/airflowServer/dags} directory, which is the working directory configured on the server for DAGs.
  \item Secure authentication: a private SSH key stored as a secret in GitHub is used, allowing automated and secure connection without exposing credentials in the code.
\end{enumerate}

This flow ensures that any modification made to workflows defined in the repository is immediately reflected in the Airflow server, allowing CI also for the orchestration layer of the MLOps system.

In Airflow, workflows are defined in Python scripts called \texttt{DAGs}, which must be stored in the directory configured as \texttt{dags\_folder} (\texttt{/home/airflow/airflowServer/dags}). 


Thanks to automatic synchronization performed from GitHub Actions via \texttt{rsync}, any change made locally to DAGs is replicated directly in that remote directory. Once there, Airflow detects changes and automatically updates workflows visible in the Web interface, from where they can be executed simply by clicking a button.

This integration allows maintaining pipeline code under version control in GitHub, while ensuring its immediate availability on the Airflow server, thus promoting DevOps best practices within the deployed MLOps system.

\paragraph{Metaflow (Sandbox)}

Unlike MLflow and Airflow, Metaflow has not been deployed on its own server, mainly due to the lack of official documentation on how to mount its user interface locally in a complete and functional manner. Instead, the environment offered by Outerbounds and Netflix known as Metaflow Sandbox was chosen to execute and validate the defined use cases.

The Metaflow sandbox is a managed, free, and cloud-accessible environment, built on Amazon SageMaker Studio Lab. This environment allows users to create Jupyter notebooks with limited access to CPU or GPU resources, without needing to configure additional infrastructure. It is accessed through \url{https://docs.outerbounds.com/sandbox/}, and when launching an environment, Metaflow is pre-installed along with its main dependencies. Once inside, the user can see various tutorials on how to perform different actions, which facilitates entry for users with less knowledge. Additionally, the sandbox interface is very similar to Visual Studio Code, making it familiar and intuitive for developers.

To execute the use cases, custom notebooks were developed that implemented Metaflow flows using the Python API. These flows were composed of functions decorated with specific instructions like \texttt{@step}, allowing structuring the training, evaluation, and visualization process in a modular way. To be able to execute them in the sandbox, it was only necessary to upload the \texttt{.py} files with the code and a \texttt{requirements.txt} file with the necessary dependencies. Results can be visualized through interactive cards automatically generated by Metaflow, which summarize relevant metrics and outputs from each execution.

Integration with the GitHub repository was performed by maintaining scripts within the corresponding subdirectory of the project. When GitHub Actions detects changes in these files, the \texttt{metaflow\_job} or \texttt{metaflow\_sentiment\_job} is executed depending on which file changes occur, which launches local execution of the scripts. Generated artifacts are automatically registered in the repository history. This execution does not directly affect the sandbox environment, as it is designed for manual development through notebooks.

Although the sandbox offers a practical solution for initial development and proofs of concept, it also presents some important limitations: it does not allow scheduled or automated flow execution, resource usage is limited, and environments expire after a certain period of inactivity. Additionally, there is no direct integration between the sandbox and GitHub Actions, so it is not possible to automate deployments or synchronization with the remote environment from repository workflows. Despite these limitations, the sandbox was useful to validate Metaflow functionality in a controlled and accessible environment without additional costs.

\paragraph{Kubeflow Pipelines (Minikube)}

Unlike MLflow and Airflow, Kubeflow Pipelines was not deployed on a cloud virtual machine, as this tool is specifically designed to run on Kubernetes environments. Instead, Minikube was used, a tool that allows creating local Kubernetes clusters on a personal machine for development and testing purposes.

Minikube was chosen as a local environment due to its ease of configuration, zero cost, and as a lightweight alternative to cloud clusters. This allows validating Kubeflow Pipelines functionality without incurring additional costs or requiring advanced cloud deployment knowledge. For installation, the steps from the official tool documentation were followed\footnote{\url{https://minikube.sigs.k8s.io/docs/start/}}.

For Kubeflow Pipelines installation, the official standalone mode procedure was followed, using Minikube with minimum resources: at least 4 CPUs and 8 GBs of memory RAM.
 
Once the cluster was started, Kubeflow Pipelines was deployed following the procedure described in its official documentation. 
Once manifests are deployed and the system is operational, port-forwarding is applied.

Pipelines designed for use cases were adapted to YAML format and executed through the Web interface that Kubeflow exposes locally (Port 8080).

YAML format scripts are integrated into the GitHub repository and, each time changes are detected in these files, the GitHub Actions workflow activates the \texttt{kubeflow\_job}, responsible for registering the change and displaying a confirmation message, since the cluster is hosted locally and is not accessible from GitHub Actions, preventing automatic file transfers.

\end{appendices}


\bibliography{bibliography}

\end{document}